\documentclass[aps,prl,twocolumn,superscriptaddress,amsmath,amssymb,floatfix]{revtex4-2}
\usepackage{graphicx}
\usepackage{bm}
\usepackage{xcolor}
\usepackage{CJK}
\usepackage[colorlinks=true,linkcolor=blue!60!black,citecolor=blue!60!black,urlcolor=blue!60!black]{hyperref}

\newcommand{\bx}{\mathbf{x}}
\newcommand{\br}{\mathbf{r}}
\newcommand{\bb}{\mathbf{b}}
\newcommand{\bd}{\bm{\delta}}
\newcommand{\bs}{\mathbf{s}}
\newcommand{\bu}{\mathbf{u}}
\newcommand{\bp}{\mathbf{p}}
\newcommand{\bphi}{\bm{\phi}}
\newcommand{\sfA}{\mathsf{A}}
\newcommand{\sfB}{\mathsf{B}}

\newcommand{\sfM}{\mathsf{M}}
\newcommand{\sfW}{\mathsf{W}}
\newcommand{\avg}[1]{\langle #1\rangle}
\newcommand{\dd}{\mathrm{d}}

\begin{document}
\raggedbottom
\begin{CJK*}{UTF8}{}
\CJKfamily{gbsn}

\title{Mutual Linearity in Nonlinear Population Dynamics}

\author{Shiling Liang (梁师翎)}
\email{shiling@pks.mpg.de}
\affiliation{Center for Systems Biology Dresden, 01307 Dresden, Germany}
\affiliation{Max Planck Institute for the Physics of Complex Systems, 01187 Dresden, Germany}
\affiliation{Max Planck Institute of Molecular Cell Biology and Genetics, 01307 Dresden, Germany}

\begin{abstract}
Mutual linearity, the persistence of affine relations among stationary observables under arbitrarily strong changes of one transition rate, is known for Markov jump processes and chemical reaction networks. We show that generalized Lotka--Volterra dynamics form a new class of mutually linear systems, even though species can go extinct and attractors can be chaotic. Any change to one species' equation acts on the others as one effective press and keeps all abundances affinely related: piecewise across extinctions, and as time averages on chaotic attractors.
\end{abstract}

\maketitle
\end{CJK*}

\textit{Introduction}---Linear response theory predicts how a stationary state shifts when a parameter is changed slightly, using only properties of the unperturbed system~\cite{Kubo1957,Marconi2008,Ruelle2009}. Many perturbations of interest are large, however, and beyond the linear regime the response of a nonlinear system generally depends on the size of the perturbation and on all parameters of the system. Exact regularities that persist in this regime are therefore rare, and valuable whenever they occur, because they make strong responses predictable.

The clearest example of such a regularity comes from Markov jump processes: when the rates of a single transition are changed by any amount, any two stationary probabilities, and any two stationary currents, remain affinely related~\cite{Harunari2024,Bebon2026,Khodabandehlou2025,Polettini2025}. This \emph{mutual linearity} complements the few other exact relations for strong responses~\cite{Bao2024,Bao2026b,Bao2026} and exact expressions and bounds for the linear response of Markov processes far from equilibrium~\cite{Owen2020,Aslyamov2024}. It has been extended to several controlled transitions~\cite{DalCengio2025}, to relaxation dynamics~\cite{Zheng2026,Voits2026}, to diffusions~\cite{Zheng2026b}, and to chemical reaction networks, where it holds for the activities of complexes~\cite{Harunari2026}. In these systems the relations come from linear stationarity conditions~\cite{Aslyamov2026}: the master equation is linear in the probabilities, and a weakly reversible, deficiency-zero chemical network has nonlinear dynamics but a stationary state fixed by linear equations for the activities. Other nonlinear dynamics with such conditions can be richer: interactions can drive components to zero, and stationary states can give way to oscillations or chaos. Whether mutual linearity survives these phenomena has remained open.

The generalized Lotka--Volterra (GLV) equations are the standard model of interacting populations in ecology and microbiology~\cite{Hofbauer1998,Stein2013} and have all of these features. Their response to strong perturbations also matters in practice: when a predator is removed, a nutrient is added, or an antibiotic is given to a gut microbiome, every species settles at a new abundance~\cite{Levins1974,Bender1984,Yodzis1988,Novak2016}. The classical theory of press perturbations predicts these abundances to first order: a sustained change of the intrinsic growth rate of species $k$, or of any other parameter that acts on $k$ alone, shifts the community along the $k$th column of the negative inverse interaction matrix~\cite{Levins1974,Bender1984,Yodzis1988,Novak2016}. In the GLV model this prediction is exact for a change of the growth rate, for holding species $k$ at a fixed abundance~\cite{Bender1984}, and for a change of one interaction coefficient~\cite{NguyenVural2022}, as long as no species goes extinct. Real perturbations are often large enough to drive species extinct, and many, such as warming, change interactions~\cite{Rall2012,Gilbert2014}. Their effects are therefore hard to predict~\cite{Yodzis1988,Novak2016}.

\begin{figure}[b]
\centering
\includegraphics[width=\columnwidth]{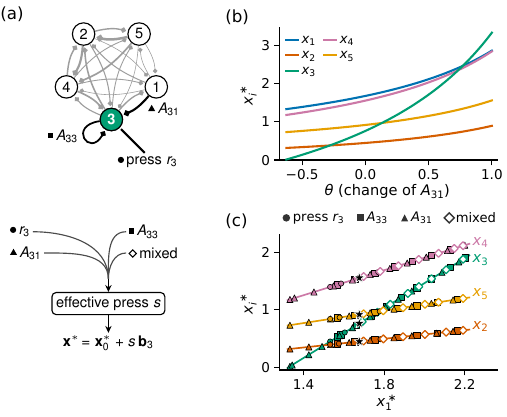}
\caption{Mutual linearity under any change to one species. (a)~A five-species community in which species~3 is modified in four ways: a press on its intrinsic growth rate $r_3$, a change of its self-regulation $A_{33}$, a change of its response $A_{31}$ to species~1, and a ``mixed'' change of several parameters at once. Each modification reaches the other species only through the effective press $s$, so the equilibrium moves along one line. (b)~When $A_{31}$ is changed to $A_{31}+\theta$, the equilibrium abundances depend nonlinearly on $\theta$. (c)~Plotted against $x_1^*$, the equilibria of all four modifications lie on the same straight lines through the unperturbed equilibrium (stars).}
\label{fig:line}
\end{figure}

\begin{figure*}[t]
\centering
\includegraphics[width=\textwidth]{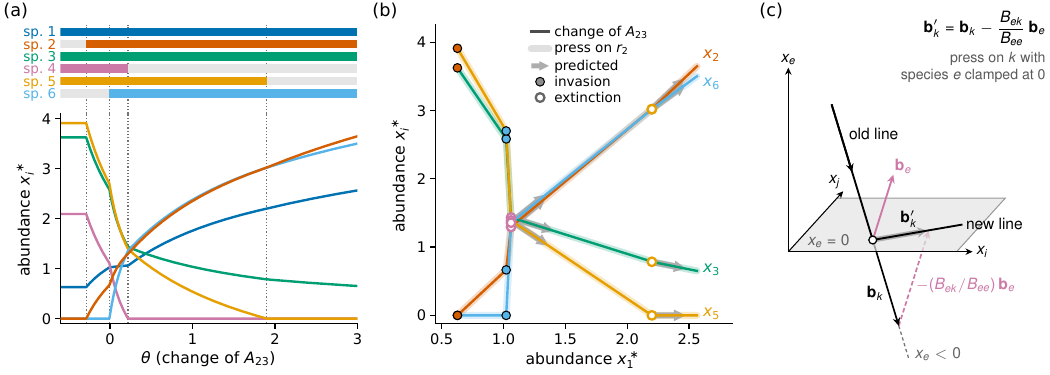}
\caption{Piecewise mutual linearity across extinctions and invasions. A six-species community in which the response $A_{23}$ of species~2 to species~3 is changed by $\theta$. (a)~Abundances against $\theta$; the strip on top shows which species are present (colored) or absent (gray). The composition changes four times (dotted lines); at the left end, species~2 itself is absent and the change has no effect. (b)~In abundance space, the path is straight between changes of composition (circles: filled where a species invades, open where one is lost, in that species' color; $x_4$ not shown) and, after each of the two extinctions, turns into the direction predicted by Eq.~\eqref{eq:kink} (gray arrows). A press $r_2\to r_2+p$ over the same range of effective press traces the same kinked path (light bands). (c)~Geometry of a kink. The equilibrium comes down the old line (direction $\bb_k$) to the plane $x_e=0$ of the lost species $e$ and continues in that plane, in the direction obtained by moving the tip of $\bb_k$ along the press response $\bb_e$ of the lost species until it reaches the plane.}
\label{fig:kink}
\end{figure*}

\begin{figure}[t]
\centering
\includegraphics[width=\columnwidth]{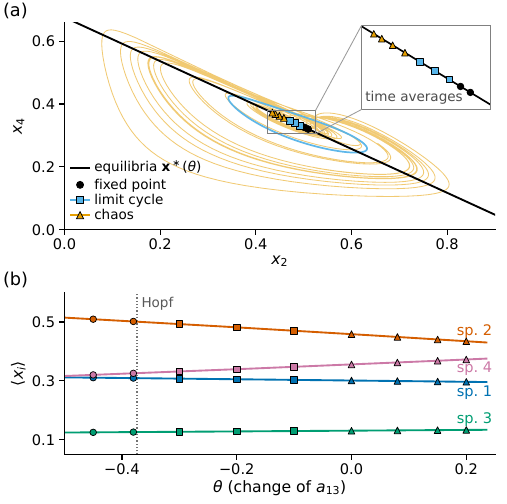}
\caption{Mutual linearity of time averages on cycles and chaos. In the competitive community of Ref.~\cite{Vano2006}, the competition coefficient $a_{13}$ of species~1 with species~3 ($A_{13}=-a_{13}$) is changed by $\theta$. (a)~A limit cycle (light blue) and a chaotic attractor (orange) in the $(x_2,x_4)$ plane, together with the line of interior equilibria $\bx^*(\theta)$ and the time averages at nine values of $\theta$ (symbols indicate the type of attractor; inset: close-up). (b)~Across a Hopf bifurcation (dotted) and the onset of chaos, the time averages follow the equilibria (lines).}
\label{fig:chaos}
\end{figure}

Here we show that GLV dynamics are mutually linear and form a new class in this family. A change to the equation of one species $k$ may alter its intrinsic growth rate, its self-regulation, or its responses to other species, in any combination and by any amount. In every case the equilibrium moves along the straight line traced by a press on $k$, so any two abundances are affinely related with coefficients independent of the control (Fig.~\ref{fig:line}). The line exists because the modification leaves the equilibrium conditions of the other species untouched, and these conditions are affine in the abundances. The other species thus respond to the modification only through a single number, an effective press, which plays the role of the current through the controlled transition of a Markov network. The relations survive extinctions, oscillations, and chaos. Unlike a probability or an activity, an abundance can reach zero. An absent species then obeys an inequality in place of its affine condition. When species go extinct or invade, mutual linearity therefore holds piecewise, with kinks in directions that we obtain in closed form and that are the same for all modifications of $k$ (Fig.~\ref{fig:kink}); on limit cycles and chaotic attractors, the time-averaged abundances obey the same relations (Fig.~\ref{fig:chaos}).

\textit{Press responses}---In the GLV model, the abundances $x_i$ of $n$ species obey
\begin{equation}
\dot x_i=x_i\,g_i(\bx),\qquad g_i(\bx)=r_i+\textstyle\sum_{j}A_{ij}x_j ,
\label{eq:glv}
\end{equation}
where $g_i$ is the per-capita growth rate of species $i$, $r_i$ its intrinsic growth rate, and $A_{ij}$ the per-capita effect of species $j$ on $i$ ($A_{ii}<0$: self-regulation)~\cite{Hofbauer1998}. Although the dynamics are nonlinear, an equilibrium at which all species coexist solves the $n$ linear equations $g_i(\bx^*)=0$, so that $\bx_0^*=\sfB\br$ with $\sfB=-\sfA^{-1}$. A \emph{press} on species $k$ is a sustained change $r_k\to r_k+p$ of its intrinsic growth rate (an added mortality has $p<0$). It therefore moves this equilibrium exactly, for any $p$, to $\sfB(\br+p\,\bd_k)=\bx_0^*+p\,\bb_k$, where $\bd_k$ is the $k$th unit vector and $\bb_k$ the $k$th column of $\sfB$~\cite{Bender1984,Yodzis1988,Novak2016}. The vector $\bb_k$ is the \emph{press response} of $k$: it sums all direct and indirect effects of $k$ on the other species.

\textit{Effective press}---Most perturbations act on more than a growth rate, so suppose now that the equation of species $k$ is changed in an arbitrary way,
\begin{equation}
\dot x_k=x_k\,[\,g_k(\bx)+h(\bx;\theta)\,],
\label{eq:mod}
\end{equation}
where $h$ is the change of $k$'s per-capita growth rate, $\theta$ stands for any number of control parameters, and $h=0$ in the unmodified community. This covers a press ($h=p$), a change of self-regulation ($h=\theta x_k$) or of the response of $k$ to another species ($h=\theta x_l$), any combination of these, and nonlinear changes such as a saturating functional response of $k$ or a constant-yield harvest ($h=-H/x_k$, for $x_k>0$). Because the equations of the other species are untouched, at a coexistence equilibrium they still read $g_i(\bx^*)=0$ for $i\neq k$, whereas species $k$ obeys $g_k(\bx^*)+s=0$ with
\begin{equation}
s:=h(\bx^*;\theta),
\label{eq:s}
\end{equation}
the change of $k$'s per-capita growth rate evaluated at the new equilibrium. These are exactly the equilibrium conditions of the \emph{unmodified} community under a press of size $s$ on species $k$, and hence
\begin{equation}
\bx^*(\theta)=\bx_0^*+s(\theta)\,\bb_k .
\label{eq:line}
\end{equation}
Whatever is done to species $k$, the community moves along the \emph{press line} of $k$, the line through $\bx_0^*$ in the direction $\bb_k$ [Fig.~\ref{fig:line}(c)]. The other species respond only to the scalar $s$, so a change of an interaction acts on them exactly as a press of size $s$ [Fig.~\ref{fig:line}(a)]; we call $s$ the \emph{effective press}. Geometrically, each unmodified species confines the equilibrium to its zero-growth hyperplane $g_i(\bx)=0$. These $n-1$ hyperplanes are independent of $\theta$, so their intersection is fixed; it is the press line, and species $k$ selects a point on it. The same hyperplanes underlie the inference of interaction networks from steady states of different subcommunities~\cite{Xiao2017}.

Mutual linearity in Markov networks has the same origin~\cite{Aslyamov2026}. When the rates of one edge change, the stationary probabilities respond only through the current across that edge, provided the network stays connected without the edge (Appendix~D). This current is the counterpart of the effective press $s$. The difference is where the linearity sits: the whole master equation is linear in the probabilities, whereas in the GLV model only the stationarity conditions of the unmodified species are linear, and these are enough for the line.

\textit{Mutual linearity}---Eliminating $s$ from two components of Eq.~\eqref{eq:line} gives the mutual-linearity relation, for any species $i$ and $j$ with $B_{jk}\neq0$,
\begin{equation}
x_i^*-x_{0,i}^*=\frac{B_{ik}}{B_{jk}}\,\big(x_j^*-x_{0,j}^*\big).
\label{eq:ML}
\end{equation}
The coefficient is a ratio of press responses and is therefore the same for every modification of $k$, whatever its size [Fig.~\ref{fig:line}(c)]; the details of the modification enter only through $s$, which Eq.~\eqref{eq:s} determines self-consistently. For a change of one interaction, $A_{kl}\to A_{kl}+\theta$, one has $s=\theta x_l^*$, while Eq.~\eqref{eq:line} gives $x_l^*=x_{0,l}^*+s\,B_{lk}$, so that
\begin{equation}
s(\theta)=\frac{\theta\,x_{0,l}^*}{1-\theta B_{lk}} .
\label{eq:mobius}
\end{equation}
Every abundance is therefore a ratio of two affine functions of $\theta$, with the same denominator for all species [Fig.~\ref{fig:line}(b)]; each stationary probability of a Markov network depends on one rate in the same way~\cite{Harunari2024}. Linear response theory gives the tangent $x_{0,l}^*\bb_k$ of this curve at $\theta=0$~\cite{Meszena2006,Barabas2014}, and Nguyen and Vural showed that this direction is exact for all $\theta$~\cite{NguyenVural2022}. By Eq.~\eqref{eq:line}, this direction is that of a press on $k$ and is therefore the same for every modification of $k$, while all the nonlinearity is in the distance $s(\theta)$ traveled along it. If $B_{lk}\neq0$, this distance diverges as $\theta$ approaches $1/B_{lk}$, but a species goes extinct first unless every entry of $\bb_k$ is zero or has the sign of $B_{lk}$ (Appendix~B).

\textit{Piecewise mutual linearity}---Equation~\eqref{eq:line} assumes that all species coexist, but a large enough change drives species extinct or lets absent ones invade [Fig.~\ref{fig:kink}(a)]. If the community starts with all species present and then settles at an equilibrium, this equilibrium is saturated~\cite{Hofbauer1998}: the per-capita growth rates of present species vanish, while absent species have $x_e^*=0$ and cannot invade, $g_e(\bx^*)\le0$. Absent species fit the same picture once this inequality is written as $g_e(\bx^*)+s_e=0$ with $s_e=-g_e(\bx^*)\ge0$: species $e$ behaves as if an effective press of its own held it at $x_e=0$, like the reaction force that holds a particle at a wall. Like that force, this press acts only at zero abundance, so $x_i^*s_i=0$ for every unmodified species $i$. With these presses, $\bx^*=\bx_0^*+s\,\bb_k+\sum_e s_e\bb_e$, and the conditions $x_e^*=0$ fix each $s_e$ as an affine function of $s$ (Appendix~A). Mutual linearity therefore holds piecewise: the path is straight as long as the composition of present species stays the same, and it kinks where the composition changes [Fig.~\ref{fig:kink}(b)]. When species $e$ is lost, the direction changes from $\bb_k$ to
\begin{equation}
\bb_k'=\bb_k-\frac{B_{ek}}{B_{ee}}\,\bb_e ,
\label{eq:kink}
\end{equation}
where now $\sfB=-\sfA_{SS}^{-1}$ for the set $S$ of species present before the extinction. The new direction is the response to a press on $k$ with $e$ clamped at zero: the part of the response that was transmitted through $e$ is removed, so the direction changes by $-(B_{ek}/B_{ee})\,\bb_e$, a correction proportional to how strongly a press on $k$ reaches $e$ [Fig.~\ref{fig:kink}(c)]. Because the lost species reaches zero abundance and zero per-capita growth rate at the same point, the two branches meet at the kink, and an invasion is the same event in reverse.

The kinks also occur at the same places for every modification of $k$. The equilibrium conditions of the other species, present or absent, are those of the unmodified community, and the condition for $k$ is that of a press $s$. Any saturated equilibrium of a modified community is therefore a saturated equilibrium of the unmodified community under the press $s$. This equilibrium is unique for every press if $-\sfA$ is a $P$-matrix (all principal minors positive)~\cite{Hofbauer1998,Cottle1992}, and it is then also globally attracting if $\sfA$ is diagonally stable ($\mathsf D\sfA+\sfA^{\mathsf T}\mathsf D$ negative definite for some positive diagonal matrix $\mathsf D$)~\cite{Takeuchi1980}. When it is unique, all modifications of $k$ move the community along one piecewise-straight curve, the \emph{press path} of $k$ [Fig.~\ref{fig:kink}(b)]. They differ only in $s(\theta)$, which sets how fast and how far the community moves along it. Classical press experiments hold species $k$ at a fixed abundance~\cite{Bender1984}. Such an experiment changes only the equation of $k$, so it also moves the other species along the press path, even beyond the first extinction. Along the path, $g_k=-s$ varies piecewise linearly with $x_k$, as found for the growth rate of invaders of Lotka--Volterra communities~\cite{Arnoldi2022}.

\textit{Time-averaged mutual linearity}---The derivation of Eq.~\eqref{eq:line} used only that the per-capita growth rates of the unmodified species vanish, and on attractors other than fixed points these rates still vanish on average. Let the species of a set $S$ persist on a bounded trajectory ($\liminf_t x_i>0$), and write $\avg{\bx}_T=T^{-1}\!\int_0^T\bx\,\dd t$ for the time average. Because $g_i$ is affine and equals $\dd\ln x_i/\dd t$, every unmodified $i\in S$ obeys
\begin{equation}
g_i(\avg{\bx}_T)=\frac1T\int_0^T g_i(\bx)\,\dd t=\frac{\ln x_i(T)-\ln x_i(0)}{T}\to0 .
\label{eq:avg}
\end{equation}
This classical argument~\cite{Volterra1931,Goel1971,Hofbauer1998} shows that time averages satisfy the equilibrium conditions of the unmodified species. The mutual-linearity relations, Eqs.~\eqref{eq:line}--\eqref{eq:kink}, hold for them with $s$ replaced by the time-averaged effective press $\avg{h}$, whether the attractor is periodic, quasiperiodic, or chaotic (Appendix~C). Figure~\ref{fig:chaos} shows this for the chaotic four-species competitive community of Vano \emph{et al.}~\cite{Vano2006}. As one competition coefficient varies, the attractor passes from a fixed point through a Hopf bifurcation to a limit cycle and to chaos, while the time averages stay on the line of (now unstable) equilibria to within the finite-time error~\cite{SM}. For nonlinear $h$ the averages still lie on the line, although $\avg{\bx}$ is in general not an equilibrium of the modified community, because $\avg{h}\neq h(\avg{\bx})$.

\textit{Generality}---The argument rests on three properties: the control acts on the equation of a single species, the per-capita growth rates of the other species are affine in the abundances, and these rates vanish, at least on average. Relaxing the first two has simple consequences. Modifying the equations of a set $K$ of species confines the response to the affine subspace $\bx_0^*+\mathrm{span}\{\bb_k\}_{k\in K}$, in which any $|K|+1$ abundances are affinely related; this is the counterpart of the multilinearity found for several controlled transitions~\cite{DalCengio2025}. A parameter that enters two equations, such as a feeding rate that is a consumer's gain and a resource's loss, therefore yields a plane, and one that enters every equation, such as temperature, can move the community in any direction. If the other species have nonlinear functional responses, their zero-growth surfaces are curved, and their intersection becomes a curve that is still shared by all modifications of $k$; nonlinearity in the equation of the modified species itself, as in Eq.~\eqref{eq:mod}, leaves the line straight. The same straight lines appear in MacArthur's consumer--resource model and in replicator dynamics, whose stationarity conditions are linear in the state (for replicators, also in the mean payoff)~\cite{SM}.

\textit{Discussion}---Mutual linearity thus holds in population dynamics: abundances can depend nonlinearly on the control, yet any two of them stay affinely related, piecewise across extinctions and for time averages on chaotic attractors. Like the known members of the family, the new class has the structure of Appendix~D: stationarity conditions that are linear in suitable variables, and a control that enters them along one fixed direction~\cite{Aslyamov2026}. In Markov and deficiency-zero chemical reaction networks the variables are probabilities or activities of complexes~\cite{Harunari2026}, and the scalar that carries the control is a current or a reaction flux; in population dynamics the variables are abundances and the scalar is the effective press. Population dynamics adds three features to this structure: controls that are themselves nonlinear, boundaries at which mutual linearity becomes piecewise [Eq.~\eqref{eq:kink}], and nonstationary attractors whose time averages obey the relations. At these boundaries inequalities replace linear conditions, so mutual linearity extends from linear equations to linear complementarity problems (Appendix~D). The path stays connected as long as the saturated equilibrium is unique. In the language of structural stability, the kinks are where the effective growth-rate vector $\br+s\,\bd_k$ crosses the boundary of a feasibility cone~\cite{Rohr2014,Saavedra2017}. Conversely, the ecological picture suggests viewing the current through a controlled transition of a Markov network as a press on the rest of the network.

Mutual linearity also makes strong responses predictable. As long as the composition stays the same, the response of one species to any manipulation of species $k$ predicts the responses of all others; synthetic microbial communities allow a direct test~\cite{Venturelli2018}. The coefficients of these relations are ratios of press responses, which are hard to compute from estimated interactions~\cite{Yodzis1988,Novak2016}. Because two points fix a line, however, two equilibrium measurements determine the mutual-linearity relations without inferring the interaction matrix.

\begin{acknowledgments}
\textit{Acknowledgments}---The author thanks K.~Katayama and Z.~Hu for pointing out Ref.~\cite{NguyenVural2022} and acknowledges financial support from the Max Planck Society.
\end{acknowledgments}

\textit{Author contributions and use of AI}---Inspired by the recent finding of mutual linearity in chemical reaction networks~\cite{Harunari2026}, whose dynamics are nonlinear, the author proposed exploring mutual linearity in other nonlinear systems. Guided by the intuition that the perturbation should be local and the system a relatively simple nonlinear one, the author chose the generalized Lotka--Volterra model as the system to study. The author also proposed examining time-averaged quantities on attractors other than fixed points, and directed the work and its presentation. Claude Opus~5.5 (Anthropic) first found mutual linearity in numerical experiments on the GLV model and then derived the results under the author's guidance. It also wrote the code that generates all data and figures, and wrote the first draft. Other Claude instances and OpenAI Codex (gpt-6-astra) first checked the results, and the author then checked and reproduced them independently. The author takes full responsibility for the Letter.

\textit{Data availability}---The data that support the findings of this Letter were generated by numerical simulations. The source code, the parameters, and the generated data are openly available~\cite{LiangCode}.

%

\section*{End Matter}
\newcommand{\appeq}[1]{\setcounter{equation}{0}\renewcommand{\theequation}{#1\arabic{equation}}\renewcommand{\theHequation}{#1\arabic{equation}}}
\appeq{A}

\textit{Appendix A: Effective presses, kinks, and the press path}---Let $\bx^*$ be a saturated equilibrium of the modified community, Eqs.~\eqref{eq:glv} and~\eqref{eq:mod}, at which species $k$ is present (if $k$ is absent, $\bx^*$ does not depend on $\theta$ as long as $k$ cannot invade). Every other species $i$ is either present, with $g_i(\bx^*)=0$, or absent, with $x_i^*=0$ and $g_i(\bx^*)\le0$. To treat both cases alike, assign to every species an effective press $s_i$: $s_k=h(\bx^*;\theta)$, $s_e=-g_e(\bx^*)\ge0$ for every absent species $e$, and $s_i=0$ for the other present species. Then $g_i(\bx^*)+s_i=0$ for all $i$, so $\sfA\bx^*+\br+\bs=0$, or
\begin{equation}
\bx^*=\bx_0^*+s_k\bb_k+\sum_{e\in E}s_e\bb_e,\qquad x_e^*=0\ \ (e\in E),
\label{eq:A1}
\end{equation}
where $E$ is the set of absent species and $\bx_0^*=\sfB\br$, $\sfB=-\sfA^{-1}$, refer to the whole community ($\bx_0^*$ may have negative entries; we assume, as is generic, that $\sfA$ and its principal submatrices are invertible). Equation~\eqref{eq:A1} restates the equilibrium conditions, and the results on lines, kinks, and the press path follow from it.

(i)~If all species are present, $E=\varnothing$ and Eq.~\eqref{eq:A1} reduces to Eq.~\eqref{eq:line}; if the equations of several species $k\in K$ are modified, each contributes a term $s_k\bb_k$.

(ii)~Write $\sfB_{EE}$ for the block of $\sfB$ with rows and columns in $E$, $\sfB_{\cdot E}$ for its columns in $E$, $\sfB_{Ek}$ for the entries $B_{ek}$ with $e\in E$, and $\bx^*_{0,E}$ for the entries of $\bx_0^*$ in $E$. On a branch with absent set $E$, the conditions $\bx_E^*=0$ give $\bs_E=-\sfB_{EE}^{-1}(\bx^*_{0,E}+s_k\sfB_{Ek})$, and Eq.~\eqref{eq:A1} becomes the straight line
\begin{equation}
\bx^*=\hat\bx_0+s_k\big(\bb_k-\sfB_{\cdot E}\,\sfB_{EE}^{-1}\sfB_{Ek}\big),
\label{eq:A2}
\end{equation}
where $\hat\bx_0=\bx_0^*-\sfB_{\cdot E}\sfB_{EE}^{-1}\bx^*_{0,E}$ is the (possibly infeasible) coexistence equilibrium of the community without $E$. The direction of this line is the response to a press on $k$ with all absent species clamped at zero; $\sfB_{EE}$ is invertible because, by Jacobi's identity, $\det\sfB_{EE}=\det(-\sfA_{SS})/\det(-\sfA)$, with $S$ the present species. For a single extinction, applying Eq.~\eqref{eq:A2} to the community of the species $S$ present before it, with $E=\{e\}$ and $\sfB=-\sfA_{SS}^{-1}$, gives Eq.~\eqref{eq:kink}. This is the update of an inverse matrix when one row and column are removed, the step that principal pivoting performs in linear complementarity problems~\cite{Cottle1992}. The result can also be checked directly: $\bb_k'$ has zero $e$ component, and $\sfA_{SS}\bb_k'=-\bd_k+(B_{ek}/B_{ee})\bd_e$, so on the remaining species $S'=S\setminus\{e\}$ it coincides with the $k$th column of $-\sfA_{S'S'}^{-1}$, the press response of the reduced community. Geometrically, $\mathrm{span}\{\bb_k,\bb_e\}$ is the plane along which the per-capita growth rates of all species but $k$ and $e$ stay fixed, and $\bb_k'$ is the direction in this plane that has zero $e$ component [Fig.~\ref{fig:kink}(c)].

(iii)~At the extinction, $x_e^*=0$ and $s_e=0$ hold simultaneously, so the new branch starts where the old one ends. An invasion is the reverse event, in which $s_e$ reaches zero on a branch where $e$ is absent.

(iv)~The conditions $g_i(\bx^*)+s_i=0$ for $i\neq k$ and the sign conditions on absent species are those of the unmodified community, so $\bx^*(\theta)$ is a saturated equilibrium of the unmodified community under the press $p=s_k(\theta)$. If that equilibrium, $\bx^*_{\mathrm{press}}(p)$, is unique for every $p$, then $\bx^*(\theta)=\bx^*_{\mathrm{press}}(s_k(\theta))$, and all modifications of $k$ move along the press path $p\mapsto\bx^*_{\mathrm{press}}(p)$, which for a $P$-matrix $-\sfA$ is continuous and piecewise linear in $p$~\cite{Cottle1992}. Each saturated equilibrium of the modified community then corresponds to a solution of $p=h(\bx^*_{\mathrm{press}}(p);\theta)$; because $\bx^*_{\mathrm{press}}$ is only piecewise linear, there can be several even for $h=\theta x_l$, all on the same path. In the community of Fig.~\ref{fig:kink}, changes of $A_{23}$, $A_{22}$, $A_{21}$, and a mixed change follow the press path of species~2 to numerical precision~\cite{SM}.

\appeq{B}\textit{Appendix B: Change of one interaction and divergence}---For $h=\theta x_l$ and a feasible $\bx_0^*$, Eq.~\eqref{eq:mobius} gives
\begin{equation}
\bx^*(\theta)=\bx_0^*+\frac{\theta\,x_{0,l}^*}{1-\theta B_{lk}}\,\bb_k ,
\label{eq:B1}
\end{equation}
the Sherman--Morrison formula for the rank-one update $\sfA+\theta\bd_k\bd_l^{\mathsf T}$~\cite{NguyenVural2022}. If $B_{lk}\neq0$, this matrix becomes singular at $\theta_\infty=1/B_{lk}$. As $\theta\to\pm\infty$, $s\to-x^*_{0,l}/B_{lk}$ and $x_l^*\to0$. Since $x_l^*=x_{0,l}^*/(1-\theta B_{lk})$ is positive only on the side of $\theta_\infty$ that contains $\theta=0$, a feasible branch can approach $\theta_\infty$ only from that side, where $s$ increases monotonically [$\dd s/\dd\theta=x_{0,l}^*/(1-\theta B_{lk})^2>0$] and diverges with the sign of $B_{lk}$. Species $i$ reaches zero along the line at $s=-x_{0,i}^*/B_{ik}$, so an extinction precedes the divergence unless every $B_{ik}$ is zero or has the sign of $B_{lk}$. Divergence without extinction therefore requires a press on $k$ to move every species in the same direction or not at all, as in the unbounded growth of Lotka--Volterra mutualists~\cite{Hofbauer1998}. In random communities, this sign criterion accounts for the observed divergences~\cite{SM}.

\appeq{C}\textit{Appendix C: Time averages}---Let $\bx(t)$ be a bounded solution on which the species of a set $S$ persist ($\liminf_{t\to\infty} x_i>0$) while the others vanish, with $k\in S$. Equation~\eqref{eq:avg} then holds for every unmodified $i\in S$. In addition, $\avg{x_e}_T\to0$ for $e\notin S$, so every limit point of $\avg{\bx}_T$ satisfies the conditions that define the line of composition $S$, Eq.~\eqref{eq:A2}, and Eq.~\eqref{eq:avg} gives the residuals of these conditions exactly; they are $O(1/T)$. The argument covers periodic, quasiperiodic, and chaotic attractors alike, but not trajectories that approach heteroclinic cycles, on which $\liminf x_i=0$.

\appeq{D}\textit{Appendix D: One structure, three systems}---Suppose that a stationary state obeys
\begin{equation}
\sfM_0\,\bphi+J\,\bu=0 ,
\label{eq:D1}
\end{equation}
a linear equation in suitable variables $\bphi\in\mathbb R^d$ with a fixed matrix $\sfM_0$ and a fixed vector $\bu$. The control enters only through the scalar amplitude $J$, which may depend on $\bphi$ and on the control parameters in any way. Then $\bphi$ lies in $V=\{\bphi:\sfM_0\bphi\in\mathrm{span}\,\bu\}$, which is independent of the control and has dimension $\dim\ker\sfM_0+1$ if $\bu\in\mathrm{range}\,\sfM_0$; each further controlled direction adds one dimension~\cite{DalCengio2025}. The same elimination of a common scalar, applied to linear equations, underlies a recent unified derivation of the Markov and chemical results from classical finite-response theory~\cite{Aslyamov2026}.
(i)~For an ecological community, $\bphi=(\bx,1)$, $\sfM_0=(\sfA\,|\,\br)$, $\bu=\bd_k$, and $J=s$. Since $\ker\sfM_0$ is spanned by $(\bx_0^*,1)$, $\dim V=2$, and fixing the last component to $1$ leaves the line~\eqref{eq:line}.
(ii)~For a Markov jump process with generator $\sfW$ and a controlled edge $a\rightleftharpoons b$ with rates $w_{ba}$ ($a\to b$) and $w_{ab}$ ($b\to a$), $\bphi=\bp$, $\sfM_0=\sfW_0$ (the generator without the edge), $\bu=\bd_b-\bd_a$ (unit vectors on the states), and $J=w_{ba}p_a-w_{ab}p_b$, the current through the edge. If the network stays connected without the edge, $\ker\sfW_0$ is spanned by its stationary state $\bm\pi_0$ and $\bu\in\mathrm{range}\,\sfW_0$, so normalization gives $\bp=\bm\pi_0+J\mathbf q$ with $\sfW_0\mathbf q=-\bu$ and $\sum_iq_i=0$. Probabilities, and the currents through all other edges, which are linear in $\bp$, are thus affine in $J$; this is the mutual linearity of Refs.~\cite{Harunari2024,Bebon2026}.
(iii)~In weakly reversible mass-action chemical networks of deficiency zero with a single linkage class, stationarity is linear in the activities of the complexes. The controlled reaction supplies $\bu$ and its flux $J$ if its removal leaves the linkage class connected. Without a linear normalization, the two-dimensional $V$ implies linear relations among any three activities, as found in Ref.~\cite{Harunari2026}.
In all three cases the controlled object acts on the rest of the system through one scalar: a current, a reaction flux, or an effective press.

Case~(i) goes beyond Eq.~\eqref{eq:D1}, because an abundance can reach zero. Equation~\eqref{eq:D1} then holds only for the species that are present, and the conditions of Appendix~A for all species read
\begin{equation}
\begin{gathered}
\sfM_0\,\bphi+J\,\bu+\textstyle\sum_{i\neq k} s_{i}\,\bd_{i}=0,\\
\phi_{i}\ge0,\qquad s_{i}\ge0,\qquad \phi_{i}s_{i}=0 ,
\end{gathered}
\label{eq:D2}
\end{equation}
with one pair $(\phi_i,s_i)$ for each unmodified species~$i$, while the modified species is present. This is a linear complementarity problem~\cite{Cottle1992} in which the control still enters only through $J$. For a fixed set of present species it reduces to Eq.~\eqref{eq:D1} for these species and gives a line, and the set changes where a $\phi_i$ or an $s_i$ reaches zero, which gives a kink. If its solution is unique for every $J$, all controls that enter along $\bu$ share one piecewise-straight path. Cases~(ii) and~(iii) are the special case in which every $\phi_i$ stays positive.

\end{document}


\begin{CJK*}{UTF8}{}
\CJKfamily{gbsn}
\title{Supplemental Material for ``Mutual Linearity in Nonlinear Population Dynamics''}
\author{Shiling Liang (梁师翎)}
\affiliation{Center for Systems Biology Dresden, 01307 Dresden, Germany}
\affiliation{Max Planck Institute for the Physics of Complex Systems, 01187 Dresden, Germany}
\affiliation{Max Planck Institute of Molecular Cell Biology and Genetics, 01307 Dresden, Germany}
\maketitle
\end{CJK*}

Equation and figure numbers without a prefix refer to the main text; those with a prefix S to this Supplemental Material.

\section{Notation}
The generalized Lotka--Volterra (GLV) model is $\dot x_i=x_ig_i(\bx)$, with per-capita growth rates $g_i(\bx)=r_i+\sum_jA_{ij}x_j$ and intrinsic growth rates $r_i$. For the whole community, $\sfB=-\sfA^{-1}$ is the press-response matrix, $\bb_k$ its $k$th column, and $\bx^*_0=\sfB\br$ the coexistence equilibrium; a press $\br\to\br+p\,\bd_k$ (with $\bd_k$ the $k$th unit vector) moves it to $\bx_0^*+p\,\bb_k$. On a set $S$ of present species the same objects are formed with $\sfB^{(S)}=-\sfA_{SS}^{-1}$ (padded with zeros for absent species). A modification of the equation of species $k$ adds $h(\bx;\theta)$ to $g_k$; its effective press is $s=h(\bx^*;\theta)$, and every coexistence equilibrium obeys $\bx^*=\bx_0^*+s\,\bb_k$ [Eq.~(4)]. With the community matrix $\sfC=\mathrm{diag}(\bx^*)\sfA$, the classical response to a press is the $k$th column of $-\sfC^{-1}$, which equals $\bb_k/x_k^*$.

\section{Numerical methods}

\textit{Random communities.} Interaction matrices have $A_{ii}=-1$ and off-diagonal entries $\gamma\,\xi_{ij}/\sqrt n$ with $\xi_{ij}$ standard normal; intrinsic growth rates $r_i$ are uniform in $[0.5,1.5]$. A draw is accepted if the interior equilibrium $\bx_0^*$ is positive ($x_{0,i}^*>0.05$) and locally stable (all eigenvalues of $\mathrm{diag}(\bx_0^*)\sfA$ with negative real parts). Fig.~1: $n=5$, $\gamma=0.9$ (seed 3); Fig.~2: $n=6$, $\gamma=1.1$ (seed 28); Fig.~\ref{fig:scope}(b): $n=7$, $\gamma=0.9$ (seed 11). The mixed modification of Fig.~1 is $r_3\to r_3+0.4\sin2\theta$, $A_{31}\to A_{31}+\theta^2$, $A_{35}\to A_{35}-0.5\theta$; that of Fig.~\ref{fig:S1}(c) is $r_2\to r_2+0.3\sin2\theta$, $A_{25}\to A_{25}+\theta^2$, $A_{26}\to A_{26}-0.7\theta$.

\textit{Equilibria along a sweep.} For each value of the control the attractor is taken to be the saturated equilibrium: among all subsets $S$ of species, the one for which $-\sfA_{SS}^{-1}\br_S$ is positive, the per-capita growth rates of the absent species are negative, and the equilibrium is locally stable. In all sweeps reported it was unique. It was checked against direct integration of the GLV equations from random initial conditions at $600$ settings, and the composition reached by integration agreed with that of the saturated equilibrium at $599$ of them; the exception is a setting one grid step from an extinction, where a species of equilibrium abundance $3\times10^{-4}$ had not yet recovered from its transient.

\textit{Checks of the line.} For the four modifications of Fig.~1 the effective press computed from its definition, Eq.~(3), agrees with the one inferred from the displacement along the line to better than $10^{-13}$, and the distance of all $304$ equilibria with full coexistence from the line $\bx_0^*+s\bb_3$ is below $10^{-13}$. The matrix of all points $(\bx^*,1)$ has numerical rank $2$ (singular values $1$, $0.29$, and then below $10^{-15}$).

\textit{Code.} Python scripts (NumPy, SciPy, Matplotlib) that generate all data and figures are openly available~\cite{LiangCode}. A script in the repository checks the numbers quoted here against their values in the text.

\section{The effective press, nonlinear functional responses, and the press path}

\begin{figure}[t]
\centering
\includegraphics[width=\textwidth]{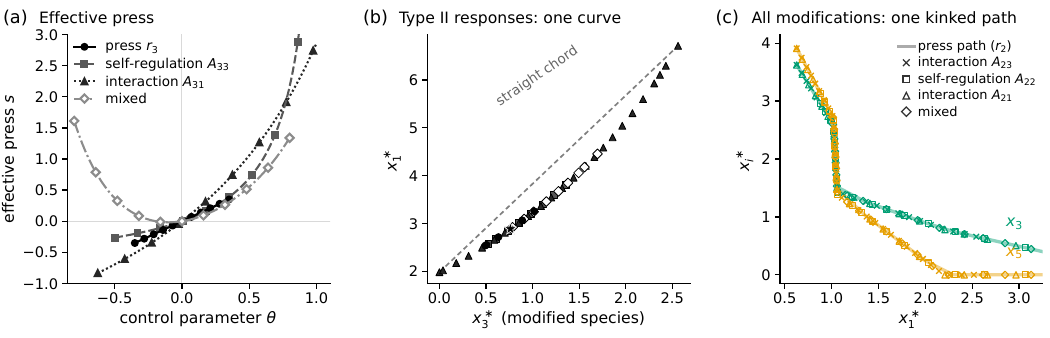}
\caption{(a)~Effective press $s$, Eq.~(3), for the four modifications of species~3 in Fig.~1: linear for a press ($s=\theta$), a ratio of affine functions for a single coefficient [Eq.~(6)], and nonmonotonic for the mixed modification. (b)~The same community with saturating (type~II) effects of species~1 and~2: the four modifications of species~3 (symbols as in a) still fall on one curve, the response curve of species~3, which is now bent (dashed: straight chord; star: unperturbed equilibrium). (c)~The community of Fig.~2: changes of $A_{23}$, $A_{22}$, $A_{21}$ and a mixed change of the equation of species~2 all move the community along the press path of species~2 (thick lines: saturated equilibria under $r_2\to r_2+p$), kinks included; shown are $x_3^*$ and $x_5^*$ against $x_1^*$.}
\label{fig:S1}
\end{figure}

\textit{Effective press.} Figure~\ref{fig:S1}(a) shows $s(\theta)$ for the modifications of Fig.~1. All the information about the type and size of a modification is in this one function; the other species respond to it as to a press.

\textit{Nonlinear functional responses.} Let $\dot x_i=x_ig_i(\bx)$ with arbitrary smooth $g_i$ and let the equation of species $k$ be modified arbitrarily. Every coexistence equilibrium then lies on $\mathcal C_k=\{\bx>0:\ g_i(\bx)=0\ \forall i\neq k\}$, which does not depend on the modification. The $n-1$ unmodified equations define $\mathcal C_k$, and the modified equation selects a point on it. By the implicit function theorem, $\mathcal C_k$ is a smooth curve wherever the $(n-1)\times n$ Jacobian of $(g_i)_{i\neq k}$ has full rank. For any two species the equilibrium abundances are therefore related by a fixed function along each branch of $\mathcal C_k$, whatever is done to $k$. For the GLV model $\mathcal C_k$ is the line $\bx_0^*+\mathrm{span}\,\bb_k$. In the type-II variant of Fig.~\ref{fig:S1}(b), in which $x_j$ is replaced by $x_j/(1+0.25\,x_j)$ for $j=1,2$ in every per-capita growth rate, and with $\mathcal C_3$ parametrized by $x_3$, the abundances obtained under the self-regulation, interaction and mixed modifications agree with those under the press to $4\times10^{-6}$ (interpolation error), while the curve deviates from a straight line by $1$--$7\%$ (relative second singular value of the centered data). Nonlinearity in the equation of the modified species itself leaves the line straight.

\textit{Universality of the press path.} Appendix~A(iv) of the main text shows that any saturated equilibrium of a modified community is a saturated equilibrium of the unmodified community under the press $p=s(\theta)$. In the community of Fig.~2 [Fig.~\ref{fig:S1}(c)], four different modifications of species~2 (changes of $A_{23}$, $A_{22}$, $A_{21}$, and a mixed change) visit five or six compositions each and agree with the press path at $p=s(\theta)$ to $10^{-13}$. The press path itself is the solution of a parametric linear complementarity problem: with $\mathsf Q=-\sfA$ and $\mathbf c=-(\br+p\,\bd_k)$, a saturated equilibrium is a solution of $\bx\ge0$, $\mathbf w=\mathsf Q\bx+\mathbf c\ge0$, $\bx\cdot\mathbf w=0$ (here $w_e=s_e$ for absent species, Appendix~A). When $\mathsf Q$ is a $P$-matrix (all principal minors positive) the solution is unique for every $p$ and continuous and piecewise linear in $p$~\cite{Cottle1992}. When several saturated equilibria coexist, the attractor can jump between branches instead (Sec.~\ref{sec:chaos}).

\section{Composition changes in random communities}

\textit{Sweeps.} In $120$ random communities ($60$ with $n=5$ and $60$ with $n=8$, $\gamma=1$), a random coefficient $A_{kl}$ was changed by $\theta\in[-3,3]$ in $121$ steps. In $93$ sweeps a saturated equilibrium existed at every step; these are used for the statistics below. The number of changes of composition per sweep had mean $1.8$ and maximum $5$; $92\%$ of the sweeps had at least one and $58\%$ at least two. Changes of composition are therefore common even when a single coefficient is changed.

\textit{Kinks.} At all $67$ events in the $93$ sweeps at which a single species was lost while $k$ and $l$ remained present (and the following branch contained at least three grid points), the direction after the kink agreed with Eq.~(7) to $1-|\cos|<10^{-15}$. In the example of Fig.~2 the third singular value of the matrix of points $(\bx^*,1)$ on each branch is about $10^{-16}$ of the first: the branches are straight.

\textit{Extinction versus divergence.} Along the line $\bx(s)=\bx_0^*+s\,\bb_k$ species $i$ reaches zero at $\hat s_i=-x_{0,i}^*/B_{ik}$. Let $\hat s_+=\min\{\hat s_i>0\}$ and $\hat s_-=\max\{\hat s_i<0\}$, with $\hat s_+=+\infty$ ($\hat s_-=-\infty$) if no species reaches zero on that side. For a change of one coefficient, $s(\theta)=\theta x_{0,l}^*/(1-\theta B_{lk})$; a feasible branch can approach the pole $\theta_\infty=1/B_{lk}$ only from the side that contains $\theta=0$, where $s$ increases monotonically and diverges with the sign of $B_{lk}$ (Appendix~B). If $s\to+\infty$ and $\hat s_+<\infty$ (or $s\to-\infty$ and $\hat s_->-\infty$), a species reaches zero before the pole: an extinction precedes the divergence. If instead every entry of $\bb_k$ is zero or has the sign of $B_{lk}$, no abundance decreases along that side of the line, no species is lost, and the equilibrium diverges at the pole. The second case occurs in purely mutualistic communities: with $\sfA=-\mathsf I+c(\mathbf 1\mathbf 1^{\mathsf T}-\mathsf I)$, $n=3$, $c=0.3$, $\br=\mathbf 1$, all entries of $\bb_1$ are positive, and increasing $A_{12}$ raises all three abundances until they diverge at $\theta_\infty=1.73$ without any extinction, the classical unbounded growth of Lotka--Volterra mutualists~\cite{Hofbauer1998}. Such divergences also occur in communities with interactions of mixed sign. In the $27$ random sweeps without a saturated equilibrium over the whole range, the last branch ran into the pole of $s(\theta)$; in $26$ of them all entries of $\bb_k$ on that branch had the sign that makes every abundance grow toward the pole, and in the remaining one an extinction and the pole lay within one grid step. In the $93$ complete sweeps, $85$ branches on which $k$ and $l$ were present ended inside the range by a change of composition, and none by a divergence.

\section{Uses of the lines}

\begin{figure}[t]
\centering
\includegraphics[width=\textwidth]{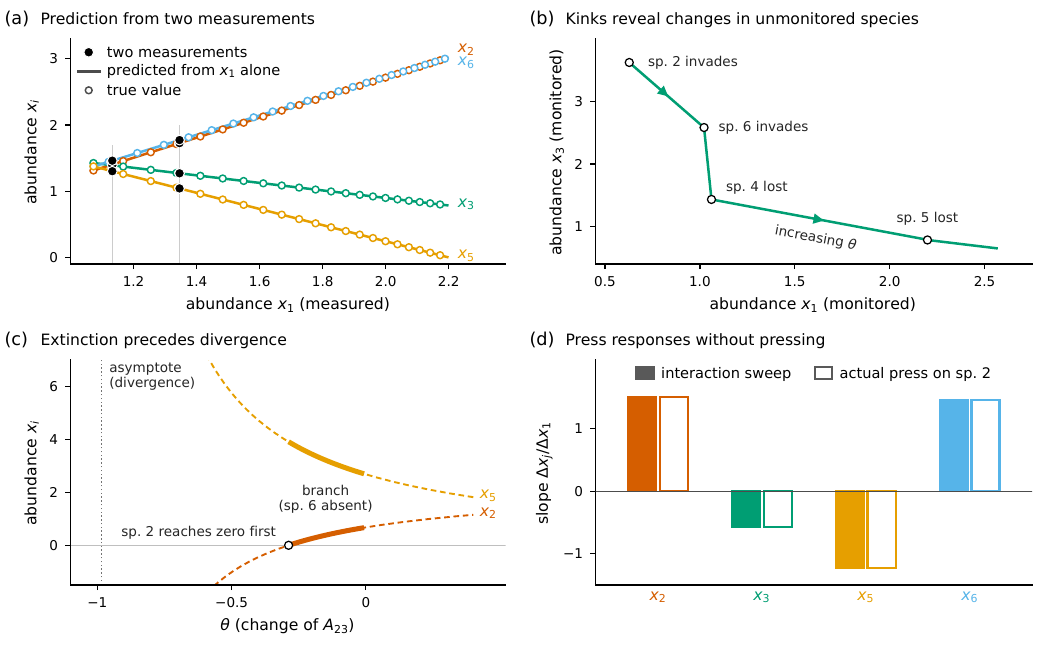}
\caption{Uses of the lines (community of Fig.~2). (a)~On the branch on which species~4 is absent, two measurements of the community at two values of $\theta$ (black) fix the line; the abundances of all species are then predicted (lines) from that of species~1 alone for every other value (open symbols: true values). (b)~Monitoring only species~1 and~3 across the whole sweep of Fig.~2: the relation between the two monitored abundances kinks at each change of composition, revealing the invasions and extinctions of unmonitored species. (c)~Along a branch each abundance is a ratio of affine functions of the control (thick: the branch on which species~6 is absent; dashed: its continuation). Moving left, the function has a pole, where the equilibrium would diverge, but species~2 reaches zero first, so the branch ends in an extinction. (d)~On the branch without species~4, the slopes of the affine relations, Eq.~(5), measured from the interaction sweep coincide with the ratios of press responses measured from an actual press on species~2. In panels (a) and (c), $x_i$ denotes equilibrium abundances $x_i^*$.}
\label{fig:uses}
\end{figure}

Figure~\ref{fig:uses} illustrates four uses of the lines. (i)~\textit{Prediction from two measurements.} The line is fixed by two points. Measuring the community at two settings of any control acting on species $k$ determines Eq.~(5) for every pair of species (measuring two species determines it for that pair) and therefore, as long as the composition does not change, predicts the response of every species from that of one species $j$ with $B_{jk}\neq0$, for every other setting of that control and for every other control acting on $k$, without knowledge of $\sfA$ [Fig.~\ref{fig:uses}(a)]. (ii)~\textit{Hidden changes of composition.} The relation between two monitored abundances generally kinks at an extinction or invasion anywhere in the community [Fig.~\ref{fig:uses}(b)]. (iii)~\textit{Press responses without pressing.} The slopes in Eq.~(5) are ratios of entries of $\bb_k$ and can be measured by manipulating an interaction of species $k$; in Fig.~\ref{fig:uses}(d), on the branch without species~4, the slopes $\Delta x_i^*/\Delta x_1^*$ from the sweep of $A_{23}$ and from an actual press on species~2 agree to all digits: $1.5146$ ($x_2$), $-0.5697$ ($x_3$), $-1.2208$ ($x_5$), and $1.4607$ ($x_6$). (iv)~\textit{Bounds.} On a branch, $s$ can range at most over the interval $(\hat s_-,\hat s_+)$ defined above, so the abundances that any manipulation of species $k$ can reach without changing the composition lie on the segment of the line between the first extinctions in either direction (an invasion may end the branch earlier); every observable affine in abundances ranges over the corresponding interval.

\section{Cycles and chaos}
\label{sec:chaos}

\begin{figure}[t]
\centering
\includegraphics[width=\textwidth]{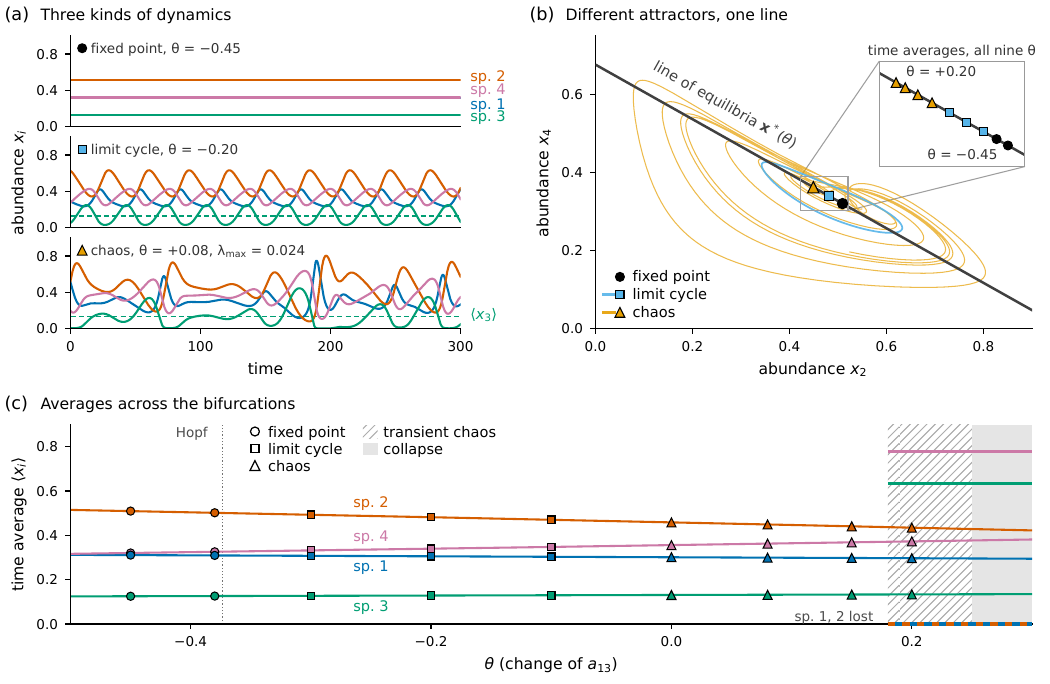}
\caption{The chaotic community of Fig.~3 in detail. The four-species competitive community of Ref.~\cite{Vano2006}, in which the competition coefficient $a_{13}$ of species~1 with species~3 ($A_{13}=-a_{13}$) is changed by $\theta$. (a)~Time series for three values of $\theta$: a stable equilibrium, a limit cycle, and a chaotic attractor (largest Lyapunov exponent $\lambda_{\max}$). The dashed line is the time average of species~3. (b)~The attractors projected on the plane of species~2 and~4 (thin lines), their time averages (symbols), and the straight line of interior equilibria $\bx^*(\theta)$; the inset zooms on the time averages, which lie on the line for all nine values of $\theta$. (c)~Time-averaged abundances against $\theta$: although the dynamics change qualitatively (dotted line: Hopf bifurcation), the averages follow the equilibria given by Eqs.~(4) and (6) (with $\theta\to-\theta$, since $A_{13}=-a_{13}$). At larger $\theta$ the chaotic attractor is destroyed: trajectories wander near it for a long transient and then escape to the equilibrium of species~3 and~4, on which the control has no effect (hatched: transient chaos or an interior attractor, depending on $\theta$ and on the initial condition; shaded: collapse from all tested initial conditions).}
\label{fig:chaos}
\end{figure}

\textit{Model and integration.} The model of Vano \emph{et al.}~\cite{Vano2006}, $\dot x_i=\rho_ix_i(1-\sum_ja_{ij}x_j)$ with $\rho=(1,0.72,1.53,1.27)$ and rows of $a$ equal to $(1,1.09,1.52,0)$, $(0,1,0.44,1.36)$, $(2.33,0,1,0.47)$, $(1.21,0.51,0.35,1)$, is of GLV form with $\br=\rho$ and $\sfA=-\mathrm{diag}(\rho)\,a$; the control $a_{13}\to a_{13}+\theta$ acts on the equation of species~1. Trajectories were integrated with an eighth-order Runge--Kutta method (relative tolerance $10^{-9}$) for a transient of $3000$ time units followed by $40\,000$ units over which the time average was taken. Lyapunov exponents were computed from the tangent dynamics over $3000$ units, with a fourth-order Runge--Kutta step of $0.01$ for the state and an Euler step for the tangent vector. The Hopf bifurcation of the interior equilibrium is at $\theta_H=-0.374$.

\begin{table}[t]
\caption{Time averages of the community of Ref.~\cite{Vano2006} under the control of $a_{13}$, over $T=4\times10^4$ time units. $\lambda_{\max}$: largest Lyapunov exponent (finite-time estimate, which changes by about $10^{-3}$ between runs; values of order $10^{-3}$ are consistent with zero, so the regime is identified from the successive maxima of $x_2$, which repeat on a cycle but not at $\theta=0.15$); $\min x$: smallest abundance on the attractor; last column: largest deviation of the time average from the interior equilibrium $\bx^*(\theta)$ on the line of Eq.~(4). On the chaotic attractors the digits of the last three columns depend on the run.}
\label{tab:chaos}
\begin{ruledtabular}
\begin{tabular}{rlrrr}
$\theta$ & regime & $\lambda_{\max}$ & $\min x$ & $\max_i|\avg{x_i}-x_i^*|$\\
\colrule
$-0.45$ & fixed point & $-0.009$ & $0.125$ & $3\times10^{-13}$\\
$-0.38$ & fixed point & $-0.001$ & $0.126$ & $3\times10^{-7}$\\
$-0.30$ & limit cycle & $0.000$ & $0.053$ & $6\times10^{-6}$\\
$-0.20$ & limit cycle & $0.000$ & $0.028$ & $1\times10^{-6}$\\
$-0.10$ & limit cycle & $-0.001$ & $0.012$ & $5\times10^{-5}$\\
$0.00$ & chaos & $0.020$ & $0.002$ & $2\times10^{-5}$\\
$0.08$ & chaos & $0.024$ & $1\times10^{-4}$ & $1\times10^{-5}$\\
$0.15$ & weak chaos & $0.003$ & $3\times10^{-8}$ & $1\times10^{-5}$\\
$0.20$ & chaos & $0.016$ & $1\times10^{-8}$ & $7\times10^{-5}$\\
\end{tabular}
\end{ruledtabular}
\end{table}

\textit{Time averages.} Table~\ref{tab:chaos} lists the time averages, and Fig.~\ref{fig:chaos} shows the attractors. Their deviations from the line are at most about $10^{-4}$ for $T=4\times10^4$, of order $1/T=2.5\times10^{-5}$, as Eq.~(8) requires; the argument does not use uniqueness, periodicity, or ergodicity of the attractor.

\textit{Compositions and the crisis.} The equilibrium of species~3 and~4, $(x_3,x_4)=(0.634,0.778)$, exists for all $\theta$ and is invaded by species~1 for $\theta<0.056$ (invasion growth rate $0.036$ at $\theta=0$, $-0.028$ at $\theta=0.1$), never by species~2. For $\theta>0.056$ the chaotic attractor and this boundary equilibrium can coexist; the smallest abundance of species~3 on the attractor falls from $2\times10^{-3}$ at $\theta=0$ to $10^{-4}$ at $\theta=0.08$, $3\times10^{-8}$ at $\theta=0.15$, and $10^{-8}$ at $\theta=0.2$. Of twelve trajectories started near the interior equilibrium at each $\theta$ from $0.15$ to $0.30$ in steps of $0.01$, and followed for $4\times10^4$ time units, all stay on the interior attractor up to $\theta=0.17$. From $\theta=0.18$ on, trajectories escape to the two-species state after a chaotic transient of $10^2$ to $3\times10^3$ time units: none stays at $\theta=0.18$ and $0.22$, between three and nine stay at $\theta=0.19$--$0.21$ and $0.23$--$0.24$, and none stays for $\theta\ge0.25$ [Fig.~\ref{fig:chaos}(c)]. The trajectory of Table~\ref{tab:chaos} at $\theta=0.2$ stays for the whole run. The path of time averages is therefore piecewise straight, but its pieces may be disconnected when several saturated states coexist.

\section{Several modified species and other models}

\begin{figure}[t]
\centering
\includegraphics[width=\textwidth]{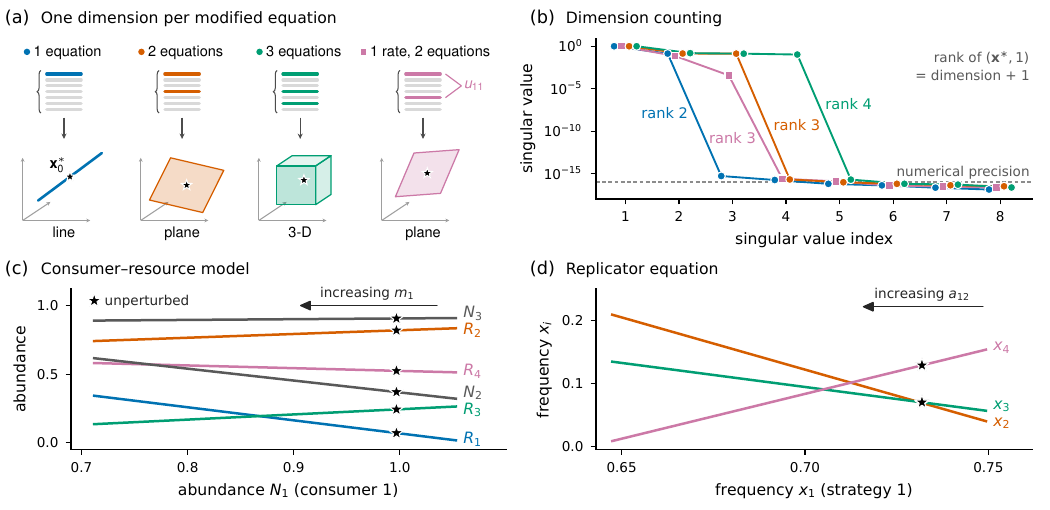}
\caption{(a)~One dimension per modified equation: modifying the equations of one, two, or three species (colored rows) confines the equilibria to a line, a plane, or a three-dimensional affine subspace through the unperturbed equilibrium $\bx^*_0$ (star). A single consumption rate shared by a consumer's and a resource's equation counts as two modified equations (right; squares in b). (b)~Singular values of the matrix of equilibria $(\bx^*,1)$ collected over random modifications of the equations of one, two, or three species of a seven-species community (circles; colors as in a; markers offset horizontally for clarity): the data span an affine subspace of dimension equal to the number of modified species (ranks 2, 3, 4; the remaining singular values are at numerical precision, dashed line). (c)~A consumer--resource community with four logistically renewed resources and three consumers: varying the mortality of consumer~1 moves all resources and consumers along straight lines. (d)~The replicator equation with four strategies: varying one payoff entry moves the interior rest point along a straight line in the simplex.}
\label{fig:scope}
\end{figure}

\textit{Several modified species.} If the equations of the species in a set $K$ are modified by functions $h_k$, every coexistence equilibrium satisfies $\bx^*=\bx_0^*+\sum_{k\in K}s_k\bb_k$ with $s_k=h_k(\bx^*;\theta)$, so the equilibria lie in the $|K|$-dimensional affine subspace $\bx_0^*+\mathrm{span}\{\bb_k\}_{k\in K}$, and any $|K|+1$ abundances obey an affine relation with control-independent coefficients. In a seven-species community, $300$ random simultaneous modifications of the intrinsic growth rate and of all interaction coefficients were drawn for each of $|K|=1,2,3$ species. The draws that keep all seven species present ($299$, $160$, and $173$) give numerical ranks $2,3,4$ of the matrix of points $(\bx^*,1)$, with the next singular value below $10^{-15}$ [Fig.~\ref{fig:scope}(b)]. A parameter that appears in two equations counts as two modified species: in the consumer--resource model below, a consumption rate $u_{11}$ shared by the equations of consumer~1 and resource~1 gives rank $3$ (singular values $1$, $0.07$, $4\times10^{-4}$, and then below $10^{-15}$). A parameter that appears in every equation, such as a temperature that changes all rates in different ways, gives no reduction at all, whereas a common rescaling of all rates leaves the equilibrium in place.

\textit{Consumer--resource model.} In the MacArthur model with four logistically renewed resources, $\dot R_j=R_j(\kappa_j-c_jR_j-\sum_iu_{ij}N_i)$, and three consumers, $\dot N_i=N_i(\epsilon_i\sum_ju_{ij}R_j-m_i)$, with random positive parameters, the per-capita growth rates are linear in the state, so the model is of GLV form on $(\mathbf R,\mathbf N)$. Varying the mortality $m_1$ of consumer~1 moves all resources and consumers along straight lines [Fig.~\ref{fig:scope}(c); numerical rank $2$ of the points $(\mathbf R,\mathbf N,1)$].

\textit{Replicator equation.} With payoff matrix $(a_{ij})$, interior rest points of $\dot x_i=x_i\big[\sum_ja_{ij}x_j-\sum_{j,l}x_ja_{jl}x_l\big]$ satisfy $\sum_ja_{ij}x_j=\lambda$ for all strategies $i$, together with $\sum_ix_i=1$. Changing one payoff entry $a_{kl}$ also changes the mean payoff in every equation, but it changes only the rest-point condition of strategy $k$, $\sum_ja_{kj}x_j=\lambda$, with $\lambda$ an unknown like the $x_i$; the rest point therefore moves along a straight line in the simplex [Fig.~\ref{fig:scope}(d): random $4\times4$ payoff matrix, entry $a_{12}$ varied; numerical rank $2$].